\def\year{2019}\relax
\documentclass[letterpaper]{article} 

\usepackage{aaai19}  
\usepackage{times}  
\usepackage{helvet}  
\usepackage{courier}  
\usepackage{url}  
\usepackage{graphicx}  
\usepackage{amssymb}
\usepackage{amsmath}
\usepackage{algorithm}
\usepackage{algorithmic}
\usepackage{amssymb}
\usepackage{multirow}
\usepackage{booktabs}
\usepackage{subfigure}
\usepackage[switch]{lineno}
\usepackage{breakurl}

\makeatletter

\newcommand{\Rmnum}[1]{\expandafter\@slowromancap\romannumeral #1@}
\makeatother

\begin{document}
%

\title{Towards Optimal Discrete Online Hashing with Balanced Similarity}
\author{Mingbao Lin$^{1}$, Rongrong Ji$^{12}\thanks{Corresponding Author.}$, Hong Liu$^{1}$, Xiaoshuai Sun$^{1}$, Yongjian Wu$^{3}$, Yunsheng Wu$^{3}$ \\ $^1$Fujian Key Laboratory of Sensing and Computing for Smart City, Department of Cognitive Science, \\ School of Information Science and Engineering, Xiamen University, China \\ $^2$ Peng Cheng Laboratory, China \\ $^3$Tencent Youtu Lab, Tencent Technology (Shanghai) Co., Ltd, China \\ lmbxmu@stu.xmu.edu.cn, rrji@xmu.edu.cn, lynnliu.xmu@gmail.com, \\ xiaoshuaisun.hit@gmail.com, \{littlekenwu, simonwu\}@tencent.com}

\maketitle
\begin{abstract}
When facing large-scale image datasets, online hashing serves as a promising solution for online retrieval and prediction tasks. It encodes the online streaming data into compact binary codes, and simultaneously updates the hash functions to renew codes of the existing dataset.
To this end, the existing methods update hash functions solely based on the new data batch, without investigating the correlation between such new data and the existing dataset.
In addition, existing works update the hash functions using a relaxation process in its corresponding approximated continuous space.
And it remains as an open problem to directly apply discrete optimizations in online hashing.
In this paper, we propose a novel supervised online hashing method, termed \textbf{B}alanced \textbf{S}imilarity for \textbf{O}nline \textbf{D}iscrete \textbf{H}ashing (BSODH), to solve the above problems in a unified framework.
BSODH employs a well-designed hashing algorithm to preserve the similarity between the streaming data and the existing dataset via an asymmetric graph regularization.
We further identify the ``data-imbalance" problem brought by the constructed asymmetric graph, which restricts the application of discrete optimization in our problem.
Therefore, a novel \emph{balanced similarity} is further proposed, which uses two equilibrium factors to balance the similar and dissimilar weights and eventually enables the usage of discrete optimizations.
Extensive experiments conducted on three widely-used benchmarks demonstrate the advantages of the proposed method over the state-of-the-art methods. 
The code is available at \url{https://github.com/lmbxmu/mycode/tree/master/2019AAAI_BSODH}.
\end{abstract}

\section{Introduction}
With the increasing amount of image data available on the Internet, hashing has been widely applied to approximate nearest neighbor (ANN) search \cite{wang2016learning,wang2018survey,lin2018supervised,liu2018dense}.
It aims at mapping real-valued image features to compact binary codes,
which merits in both low storage and efficient computation on large-scale datasets.
One promising direction is online hashing (OH), which has attracted increasing attentions recently.
Under such an application scenario, data are often fed into the system via a streaming fashion, while traditional hashing methods can hardly accommodate this configuration.
In OH, the online streaming data is encoded into compact binary codes, while the hash functions are simultaneously updated in order to renew codes of the existing data.

\begin{figure}[!t]
\begin{center}
\includegraphics[height=0.45\linewidth]{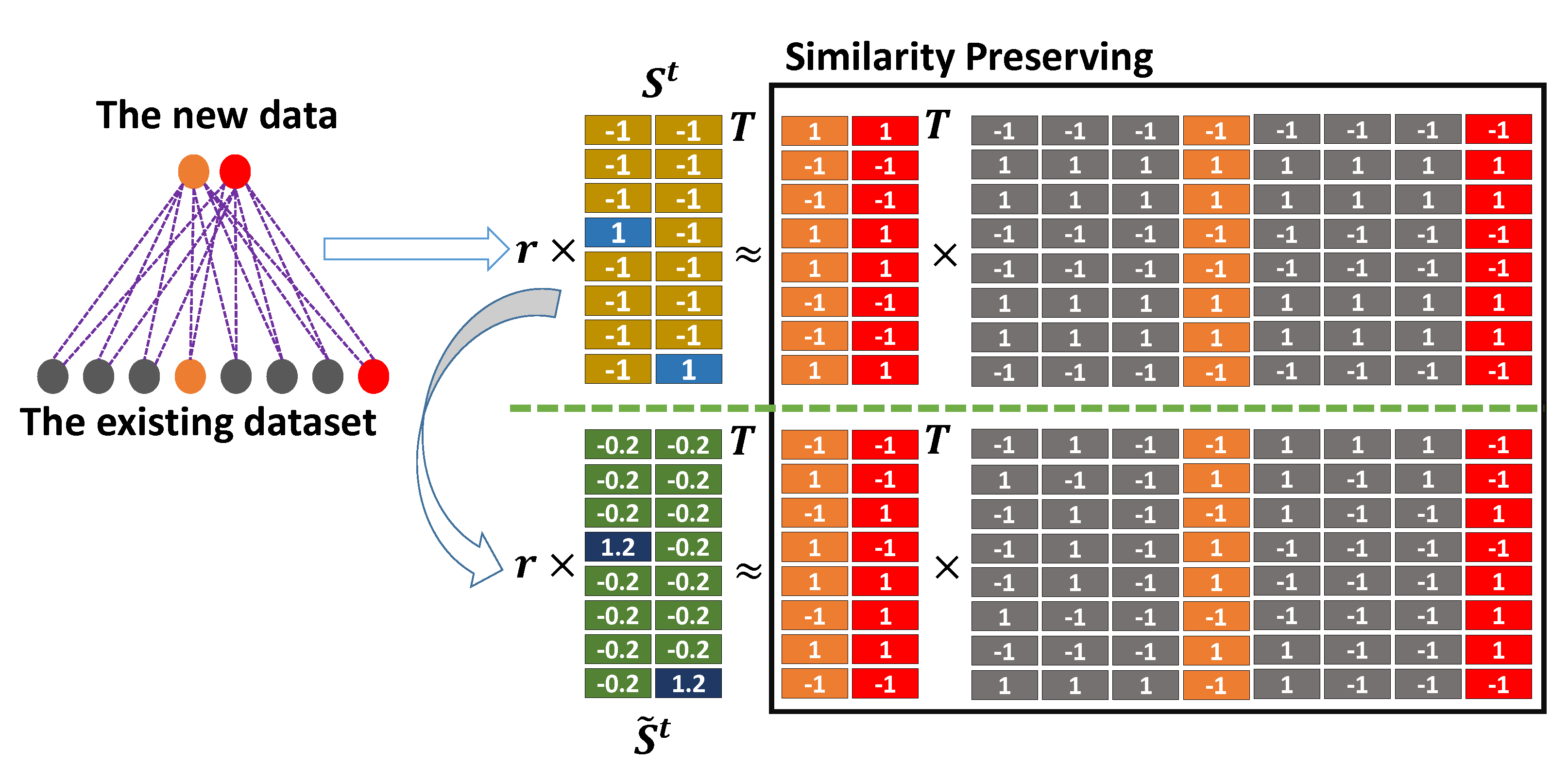}
\end{center}
\caption{\label{system}
An example of data-imbalance problem and the learned binary codes. The similarity matrix $\mathbf{S}^t$ is highly sparse under online setting and thus tends to generate consistent binary codes, which are indiscriminate and uninformative. With the introduction of the balanced similarity $\mathbf{\tilde{S}}^t$, codes of similar items are tightened while codes of dissimilar items are expanded. By combining with discrete optimizations, advanced retrieval results are obtained.
}
\end{figure}

In principle, OH aims to analyze the streaming data while preserving structure of the existing
dataset\footnote{The streaming data is usually in a small batch, which can be processed easily to pursue a better tradeoff among computation, storage, and accuracy.}.
In the literature, several recent works have been proposed to handle OH.
The representative works include, but not limited to, OKH \cite{huang2013online}, SketchHash \cite{leng2015online}, AdaptHash \cite{cakir2015adaptive}, OSH \cite{cakir2017online}, FROSH \cite{Chen2017FROSHFO} and MIHash \cite{fatih2017mihash}.
However, the performance of OH is still far from satisfactory for real-world applications.
We attribute it to two open issues, \emph{i.e.}, \emph{updating imbalance} and \emph{optimization inefficiency}.
In terms of the updating imbalance, the existing OH schemes update hash functions solely based on the newly coming data batch, without investigating the correlation between such new data and the existing dataset.
To that effect, an asymmetric graph can be constructed to preserve similarity between the new data and the existing dataset as shown in Fig.\ref{system}. Under online setting, the similarity matrix is usually sparse and unbalanced, \emph{i.e.}, data-imbalance phenomenon, since most image pairs are dissimilar and only a few are similar.
The updating imbalance issue,  if not well addressed, might cause the learned binary codes ineffective for both the new data and the existing data, and hence lead to severe performance degeneration for OH schemes.

In terms of the optimization inefficiency, the existing OH schemes still rely on the traditional relaxation \cite{gong2011iterative,datar2004locality,jiang2015scalable,liu2018ordinal,lin2018holistic} over the approximated continuous space to learn hash functions, which often makes the produced hash functions less effective, especially when the code length increases \cite{liu2014discrete,shen2015supervised}.
Despite the recent advances in direct discrete optimizations in offline hashing \cite{ji2017toward,jiang2018asymmetric} with discrete cyclic coordinate descent (DCC) \cite{shen2015supervised},
such discrete optimizations can not be directly applied to online case that contains serious data-imbalance problem, since the optimization heavily relies on the dissimilar pairs, and thus lose the information of similar pairs.

We argue that, the above two issues are not independent.
In particular, to conduct discrete optimizations, the existing offline methods typically adopt an asymmetric graph regularization to preserve the similarity between training data.
Constructing the asymmetric graph consumes both time and memory.
Note that, since the streaming data is in a small batch, such an asymmetric graph between the streaming data and the existing dataset can be dynamically constructed under online setting.
However, as verified both theoretically and experimentally later, it still can not avoid the generation of consistent codes (most bits are the same) due to the data-imbalance problem brought by the constructed asymmetric graph in online learning, as illustrated in Fig.\ref{system}.

In this paper, we propose a novel supervised OH method, termed Balanced Similarity for Online Discrete Hashing (BSODH) to handle the updating imbalance and optimization inefficiency problems in a unified framework.
First, unlike the previous OH schemes, the proposed BSODH mainly considers updating the hash functions with correlation between the online streaming data and the existing dataset.
Therefore, we aim to adopt an asymmetric graph regularization to preserve the relation in the produced Hamming space.
Second, we further integrate the discrete optimizations into OH, which essentially tackles the challenge of quantization error brought by the relaxation learning.
Finally, we present a new similarity measurement, termed \emph{balanced similarity}, to solve the problem of data-imbalance during the discrete binary learning process.
In particular, we introduce two equilibrium factors to balance the weights of similar and dissimilar data, and thus enable the discrete optimizations.
Extensive experimental results on three widely-used benchmarks, \emph{i.e.}, CIFAR$10$, Places$205$ and MNIST, demonstrate the advantages of the proposed BSODH over the state-of-the-art methods.

To summarize, the main contributions of the proposed BSODH in this paper include:
\begin{itemize}
  \item To capture the data correlation between online streaming data and the existing dataset, we introduce an asymmetric graph regularization to preserve such correlation in the produced Hamming space.
  \item To reduce the quantization error in the Hamming space, we design a customized discrete optimization algorithm. It handles the optimization inefficiency issue in the existing OH scheme, making discrete learning feasible for the first time in the online framework.
  \item We propose a balanced similarity matrix to handle the data-imbalance problem, which further prevents the generation of consistent binary codes, \emph{i.e.}, a phenomenon that previously occurred when directly applying discrete optimizations in online setting.
\end{itemize}

\section{Related Work}
In this section, we briefly review the existing OH methods.
OH merits in efficiently updating the hash functions by using the streaming data online, which can be further subdivided into two categories: SGD-based OH methods, and matrix sketch-based OH methods.

For SGD-based methods, Online Kernel Hashing (OKH) \cite{huang2013online} is the first attempt to learn hash functions via an online passive-aggressive strategy \cite{crammer2006online}, which updates hash functions to retain important information while embracing information from new pairwise input.
Adaptive Hashing (AdaptHash) \cite{cakir2015adaptive} adopts a hinge loss to decide which hash function to be updated.
Similar to OKH, labels of pairwise similarity are needed for AdaptHash.
Inspired by Error Correcting Output Codes (ECOCs) \cite{dietterich1995solving},
Online Supervised Hashing (OSH) \cite{cakir2017online} adopts a more general two-step hash learning framework,
where each class is firstly deployed with a vector from ECOCs,
and then an convex function is further exploited to replace the $0/1$ loss.
In \cite{fatih2017mihash}, an OH with Mutual Information (MIHash) is developed which targets at optimizing the mutual information between neighbors and non-neighbors.

Motivated by the idea of ``data sketching'' \cite{clarkson2009numerical},
skech-based methods provide a good alternative for unsupervised online binary coding,
via which a large dataset is summarized by a much smaller data batch.
Leng \emph{et al.} proposed the Online Sketching Hashing (SketchHash) \cite{leng2015online}, which adopts an efficient variant of SVD decomposition to learn hash functions.
More recently, Subsampled Randomized Hadamard Transform (SRHT) is adopted in FasteR Online Sketching Hashing (FROSH) \cite{Chen2017FROSHFO} to accelerate the training  process of SketchHash.

However, existing sketch-based algorithms are based on unsupervised learning, and their retrieval performance is mostly unsatisfactory without fully utilizing label information.
Although most SGD-based algorithms aim to preserve the label information via online hash function learning, the relaxation process is adopted to update the hash functions, which contradicts with the recent advances in offline hashing where discrete optimizations are adopted directly, such as Discrete Graph Hashing \cite{liu2014discrete} and Discrete Supervised Hashing \cite{shen2015supervised}.
In this paper, we are the first to investigate OH with discrete optimizations, which have shown superior performance compared with the quantization-based schemes.

\section{The Proposed Method}
\subsection{Problem Definition}
Given a dataset $\mathbf{X} = [\mathbf{x}_1, ..., \mathbf{x}_n] \in \mathbb{R}^{d \times n}$ with its corresponding labels $\mathbf{L} =  [l_1, ..., l_n] \in \mathbb{N}^{n}$, where $\mathbf{x}_i \in \mathbb{R}^d$ is the $i$-th instance with its class label $l_i \in \mathbb{N}$.
The goal of hashing is to learn a set of $k$-bit binary codes $\mathbf{B} = [\mathbf{b}_1,...,\mathbf{b}_n] \in \{-1, +1\}^{k \times n}$, where $\mathbf{b}_i$ is the binary vector of $\mathbf{x}_i$.
A widely-adopted hash function is the linear hash mapping \cite{gong2011iterative,cakir2017online}, \emph{i.e.},
\begin{equation} \label{eq1}
    \mathbf{B} = F(\mathbf{X}) = \mathrm{sgn}(\mathbf{W}^T\mathbf{X}),
\end{equation}
where $\mathbf{W} = [\mathbf{w}_1, ..., \mathbf{w}_k] \in \mathbb{R}^{d \times k}$ is the projection matrix to be learned with $\mathbf{w}_i$ being responsible for the $i$-th hash bit. The sign function $\mathrm{sgn}(x)$ returns $+1$ if input variable $x > 0$, and returns $-1$ otherwise.

For the online learning problem, the data is coming in a streaming fashion. Therefore $\textbf{X}$ is not available once for all.
Without loss of generality, we denote $\mathbf{X}_s^t = [\mathbf{x}_{s1}^t, ..., \mathbf{x}^t_{sn_t} ] \in \mathbb{R}^{d \times n_t}$ as the input streaming data at $t$-stage, and denote $\mathbf{L}_s^t = [l_{s1}^t, ..., l_{sn_t}^t] \in \mathbb{N}^{n_t}$ as the corresponding label set, where $n_t$ is the size of the batch.
We denote $\mathbf{X}_e^t = [\mathbf{X}_s^1, ..., \mathbf{X}_s^{t-1}] = [\mathbf{x}_{e1}^t, ..., \mathbf{x}_{em_t}^t] \in \mathbb{R}^{d \times m_t}$, where $m_t = n_1 + ... + n_{t-1}$, as the previously existing dataset with its label set $\mathbf{L}_e^t = [\mathbf{L}_s^1, ..., \mathbf{L}_s^{t-1}] = [l_{e1}^t,...,l_{em_t}^t] \in \mathbb{N}^{m_t}$.
Correspondingly, we denote $\mathbf{B}_s^t = \mathrm{sgn}({\mathbf{W}^t}^T\mathbf{X}_s^t) = [\mathbf{b}_{s1}^t, ..., \mathbf{b}_{sn_t}^t] \in \mathbb{R}^{k \times n_t}$, $\mathbf{B}_e^t = \mathrm{sgn}({\mathbf{W}^t}^T\mathbf{X}_e^t) = [\mathbf{b}_{e1}^t,...,\mathbf{b}_{em_t}^t] \in \mathbb{R}^{k \times m_t}$ as the discretely learned binary codes for $\mathbf{X}_s^t$ and $\mathbf{X}_e^t$, respectively.
Under online setting, the parameter matrix $\mathbf{W}^t$ should be updated based on the newly coming batch $\mathbf{X}_s^t$ instead of the existing dataset $\mathbf{X}_e^t$.

\subsection{The Proposed Framework}
Ideally, if data $\mathbf{x}_i$ and $\mathbf{x}_j$ are similar, the Hamming distance between their binary codes should be minimized, and vice versa.
This is achieved by minimizing the quantization error between the similarity matrix and the Hamming similarity matrix \cite{liu2012supervised}.
However, considering the streaming batch data alone does not reflect the structural relationship of all data samples.
Therefore, following \cite{shen2015learning,jiang2018asymmetric}, we resort to preserve the similarity in the Hamming space between new data batch $\mathbf{X}^t_s$ and the existing dataset $\mathbf{X}^t_e$ at $t$-stage with an asymmetric graph as shown in Fig.\ref{system}.
To that effect, we minimize the Frobenius norm loss between the supervised similarity and the inner products of $\mathbf{B}_s^t$ and $\mathbf{B}_e^t$  as follows:
\begin{equation} \label{original}
\begin{split}
&\min_{\mathbf{B}_s^t, \mathbf{B}_e^t} \|{\mathbf{B}_s^t}^T\mathbf{B}_e^t - k\mathbf{S}^t\|^2_{\mathcal{F}}\\ s.t. \quad \mathbf{B}_s^t \in &\{-1, 1\}^{k \times n_t}, \mathbf{B}_e^t \in \{-1, 1\}^{k \times m_t}.
\end{split}
\end{equation}
where $\mathbf{S}^t \in \mathbb{R}^{n_t \times m_t}$ is the similarity matrix between $\mathbf{X}_s^t$ and $\mathbf{X}_e^t$. Note that $s^t_{ij} = 1$ iff both $\mathbf{x}_{si}^t$ and $\mathbf{x}_{ej}^t$ share the same label, \emph{i.e.}, $l_{si}^t = l_{ej}^t$.
Otherwise, $s_{ij}^t = -1$\footnote{At each stage, $\mathbf{S}^t$ is calculated on-the-fly.}. And $\|\cdot\|_{\mathcal{F}}$ denotes the Frobenius norm.

Besides, we aim to learn the hash functions by minimizing the error term between the linear hash functions $F$ in Eq.\ref{eq1} and the corresponding binary codes $\mathbf{B}^t_s$, which is constrained by $\|\mathbf{B}^t_s - F(\mathbf{X}^t_s)\|^2_{\mathcal{F}}$.
It can be easily combined with the above asymmetric graph that can be seen as a regularizer for learning the hash functions, which is rewritten as:
\begin{equation} \label{original_binary_independent}
\begin{split}
&\min_{\mathbf{B}_s^t, \mathbf{B}_e^t,{} \mathbf{W}^t} \underbrace{\|{\mathbf{B}_s^t}^T\mathbf{B}_e^t - k\mathbf{S}^t\|^2_{\mathcal{F}}}_{\text{term 1}} + {\sigma}^t\underbrace{\|F(\mathbf{X}_s^t) - \mathbf{B}_s^t\|^2_{\mathcal{F}}}_{\text{term 2}}+ \\& {\lambda}^t\underbrace{\|\mathbf{W}^t\|^2_{\mathcal{F}}}_{\text{term 3}}
\quad s.t. \, \mathbf{B}_s^t \in \{-1, 1\}^{k \times n_t}, \mathbf{B}_e^t \in \{-1, 1\}^{k \times m_t},
\end{split}
\end{equation}
where ${\sigma}^t$ and ${\lambda}^t$ serve as two constants at $t$-stage to balance the trade-offs among the three learning parts.

We analyze that using such a framework can learn better coding functions.
Firstly, in term $2$, $\mathbf{W}^t$ is optimized based on the dynamic streaming data $\mathbf{X}_s^t$, which makes the hash function more adaptive to unseen data.
Secondly, As in Eq.\ref{w_solution}, the training complexity for $\mathbf{X}_s^t$-based learning $\mathbf{W}^t$ is $\mathcal{O}(d^2n_t + d^3)$, while it is $\mathcal{O}(d^2m_t + d^3)$ for the learnt $\mathbf{W}^t$ based on $\mathbf{X}_e^t$. Therefore, updating $\mathbf{W}^t$ based on $\mathbf{X}_e^t$ is impractical when $m_t \gg n_t$ with the increasing number of new data batch. Further, it also violates the basic principle of OH that $\mathbf{W}^t$ can only be updated based on the newly coming data.
Last but not least, with the asymmetric graph loss in term $1$, the structural relationship in the original space can be well preserved in the produced Hamming space, which makes the learned binary codes $\mathbf{B}^t_s$ more robust.
The above discussion will be verified in the subsequent experiments.

\subsection{The Data-Imbalance Issue}
As shown in Fig.\ref{system}, the similarity matrix $\mathbf{S}^t$ between the streaming data and the existing dataset is very sparse\footnote{Here, ``sparse'' denotes the vast majority of elements in a matrix are $-1$.}. That is to say, there exists a severe data-imbalance phenomenon, \emph{i.e.}, most of image pairs are dissimilar and few pairs are similar.
Due to this problem, the optimization will heavily rely on the dissimilar information and miss the similar information, which leads to performance degeneration.

As a theoretical analysis, we decouple the whole sparse similarity matrix into two subparts, where similar pairs and dissimilar pairs are separately considered. Term $1$ in Eq.\ref{original_binary_independent} is then reformulated as:
\begin{equation} \label{imbalance}
\begin{split}
&\text{term 1} \!\! =\!\! \underbrace{\sum_{i,j,\mathbf{S}_{ij}^t = 1}({{}\mathbf{b}_{si}^t}^T\mathbf{b}_{ej}^t - k)^2}_{\text{term} \; \mathcal{A}} + \!\!
\underbrace{\sum_{i, j, \mathbf{S}_{ij}^t = -1}({{}\mathbf{b}_{si}^t}^T\mathbf{b}_{ej}^t + k)^2}_{\text{term} \; \mathcal{B}}
\\& \qquad \qquad s.t. \quad {\mathbf{b}}^t_{si} \in \{-1, 1\}^k, {\mathbf{b}}^t_{ej} \in \{-1,1\}^k.
\end{split}
\end{equation}

\textbf{Analysis 1}.
We denote $\mathbf{S}^t_1 = \{ \mathbf{S}^t_{ij} \in \mathbf{S}^t | \mathbf{S}^t_{ij} = 1 \}$, $\emph{i.e.}$, the set of similar pairs and $\mathbf{S}^t_2 = \{ \mathbf{S}^t_{ij} \in \mathbf{S}^t | \mathbf{S}^t_{ij} = -1 \}$, $\emph{i.e.}$, the set of dissimilar pairs.
In online setting, when $n_t \ll m_t$ with the increase of new data batch,
the similarity matrix $\mathbf{S}^t$ becomes a highly sparse matrix, \emph{i.e.}, $|\mathbf{S}^t_1| \ll |\mathbf{S}^t_2|$.
In other words, term $1$ suffers from a severe data-imbalance problem. Furthermore, since term $1$ $\gg$ term $2$ in Eq.\ref{original_binary_independent} and term $\mathcal{B}$ $\gg$ term $\mathcal{A}$ in Eq.\ref{imbalance}, the learning process of $\mathbf{B}^t_s$ and $\mathbf{B}^t_e$ heavily relies on term $\mathcal{B}$.

A suitable way to minimize term $\mathcal{B}$ is to have ${{}\mathbf{b}^t_{si}}^T\mathbf{b}^t_{ej} = -k$, $\emph{i.e.}$, $\mathbf{b}^t_{si} = -\mathbf{b}^t_{ej}$.
Similarly, for any $\mathbf{b}^t_{eg} \in \mathbf{B}_e^t$ with $g \neq j$, we have $\mathbf{b}^t_{si} = -\mathbf{b}^t_{eg}$.
It is easy to see that $\mathbf{b}^t_{ej} = \mathbf{b}^t_{eg}$.
In other words, each item in $\mathbf{B}_e^t$ shares consistent binary codes.
Similarly, each item in $\mathbf{B}_s^t$ also shares consistent binary codes which are opposite with $\mathbf{B}_e^t$.
Fig.\ref{system} illustrates such an extreme circumstance.
However, as can be seen from term $2$ in Eq.\ref{original_binary_independent}, the performance of hash functions deeply relies on the learned $\mathbf{B}_s^t$.
Therefore, such a data-imbalance problem will cause all the codes produced by $\mathbf{W}^t$ to be biased, which will seriously affect the retrieval performance.
\
\subsection{Balanced Similarity}
To solve the above problem, a common method is to keep a balance between term $1$ and term $2$ in Eq.\ref{original_binary_independent} by scaling up the parameter ${\sigma}^t$.
However, as verified later in our experiments (see Fig.\ref{sigma}), such a scheme still suffers from unsatisfactory performance and will get stuck in how to choose an appropriate value of ${\sigma}^t$ from a large range\footnote{Under the balanced similarity, we constrain ${\sigma}^t$ to [0, 1].}.
Therefore, we present another scheme to handle this problem, which expands the feasible solutions for both $\mathbf{B}_e^t$ and $\mathbf{B}_s^t$.
Concretely, we propose to use a balanced similarity matrix $\mathbf{\tilde{S}}^t$ with each element defined as follows:
\begin{equation} \label{Softened_similarity}
\mathbf{\tilde{S}}^t_{ij} =
\begin{cases}
{\eta}_s\mathbf{S}^t_{ij}, & \mathbf{S}^t_{ij} = 1, \\
{\eta}_d\mathbf{S}^t_{ij}, & \mathbf{S}^t_{ij} = -1,
\end{cases}
\end{equation}
where ${\eta}_s$ and ${\eta}_d$ are two positive equilibrium factors used to balance the similar and dissimilar weights, respectively.
When setting ${\eta}_s > {\eta}_d$, the Hamming distances among similar pairs will be reduced, while the ones among dissimilar pairs will be enlarged.

\textbf{Analysis 2}. With the balanced similarity, the goal of term $\mathcal{B}$ in Eq.\ref{imbalance} is to have ${{}\mathbf{b}^t_{si}}^T\mathbf{b}^t_{ej} \approx -k{\eta}_{d}$.
The number of common hash bits between $\mathbf{b}^t_{si}$ and $\mathbf{b}^t_{ej}$ is at least $\lfloor\frac{k(1-{\eta}_d)}{2}\rfloor$\footnote{$\lfloor \cdot \rfloor$ denotes the operation of rounding down.}.
Therefore, by fixing $\mathbf{b}^t_{si}$, the cardinal number of feasible solutions for $\mathbf{b}^t_{ej}$ is at least $\binom{k}{\lfloor\frac{k(1-{\eta}_d)}{2}\rfloor}$.
Thus, the balanced similarity matrix $\mathbf{\tilde{S}}^t$ can effectively solve the problem of generating consistent binary codes, as showed in Fig.\ref{system}.
By replacing the similarity matrix $\mathbf{S}^t$ in Eq.\ref{original_binary_independent} with the balanced similarity matrix $\mathbf{\tilde{S}}^t$, the overall objective function can be written as:
\begin{equation} \label{original_binary_final}
\begin{split}
&\min_{\mathbf{B}_s^t, \mathbf{B}_e^t,{} \mathbf{W}^t} \underbrace{\|{\mathbf{B}_s^t}^T\mathbf{B}_e^t - k\mathbf{\tilde{S}}^t\|^2_{\mathcal{F}}}_{\text{term 1}} + {\sigma}^t\underbrace{\|F(\mathbf{X}_s^t) - \mathbf{B}_s^t\|^2_{\mathcal{F}}}_{\text{term 2}}+ \\& {\lambda}^t\underbrace{\|\mathbf{W}^t\|^2_{\mathcal{F}}}_{\text{term 3}}
\quad s.t. \, \mathbf{B}_s^t \in \{-1, 1\}^{k \times n_t}, \mathbf{B}_e^t \in \{-1, 1\}^{k \times m_t}.
\end{split}
\end{equation}

\subsection{The Optimization}
Due to the binary constraints, the optimization problem of Eq.\ref{original_binary_final} is still non-convex with respect to $\mathbf{W}^t, \mathbf{B}_s^t, \mathbf{B}_e^t$.
To find a feasible solution, we adopt an alternative optimization approach, \emph{i.e.}, updating one variable with the rest two fixed until convergence.

\textbf{1) $\mathbf{W}^t$-step}: Fix $\mathbf{B}_e^t$ and $\mathbf{B}_s^t$, then learn hash weights $\mathbf{W}^t$.
This sub-optimization of Eq.\ref{original_binary_final} is a classical linear regression that aims to find the best projection coefficient $\mathbf{W}^t$ by minimizing term 2 and term 3 jointly. Therefore, we update $\mathbf{W}^t$ with a close-formed solution as:
\begin{equation} \label{w_solution}
\mathbf{W}^t = {\sigma}^t({\sigma}^t\mathbf{X}_s^t{\mathbf{X}_s^t}^T +
{\lambda}^t\mathbf{I})^{-1}\mathbf{X}_s^t{\mathbf{B}_s^t}^T,
\end{equation}
where $\mathbf{I}$ is a $d \times d$ identity matrix.

\textbf{2) $\mathbf{B}_e^t$-step}: Fix $\mathbf{W}^t$ and $\mathbf{B}_s^t$, then update $\mathbf{B}_e^t$. Since only term $1$ in Eq.\ref{original_binary_final} contains $\mathbf{B}_e^t$, we directly optimize this term via a discrete optimization similar to \cite{kang2016column}, where the squared Frobenius norm in term 1 is replaced with the $L_1$ norm. The new formulation is:
\begin{equation} \label{be_l1}
\min_{\mathbf{B}_e^t} \|{\mathbf{B}_s^t}^T\mathbf{B}_e^t - k\mathbf{\tilde{S}}^t\|_1
\quad s.t. \quad \mathbf{B}_e^t \in \{-1, 1\}^{k \times m_t}.
\end{equation}

Similar to \cite{kang2016column}, the solution of Eq.\ref{be_l1} is as follows:
\begin{equation} \label{be_solution}
\mathbf{B}_e^t = sgn(\mathbf{B}_s^t\mathbf{\tilde{S}}^t).
\end{equation}

\textbf{3) $\mathbf{B}_s^t$-step}: Fix $\mathbf{B}_e^t$ and $\mathbf{W}^t$, then update $\mathbf{B}_s^t$. The corresponding sub-problem is:
\begin{equation} \label{eq_bs}
\begin{split}
&\min_{\mathbf{B}_s^t}
\|{\mathbf{B}_s^t}^T\mathbf{B}_e^t - k\mathbf{\tilde{S}}^t\|^2_{\mathcal{F}}
+ {\sigma}^t\|{\mathbf{W}^t}^T\mathbf{X}_s^t - \mathbf{B}_s^t\|^2_{\mathcal{F}}
\\ & \qquad\qquad\quad s.t. \quad \mathbf{B}_s^t \in \{-1, 1\}^{k \times n_t}.
\end{split}
\end{equation}

By expanding each term in Eq.\ref{eq_bs}, we get the sub-optimal problem of $\mathbf{B}_s^t$ by minimizing the following formulation:
\begin{equation} \label{expand}
\begin{split}
&\min_{\mathbf{B}_s^t}\| {\mathbf{B}_e^t}^T\mathbf{B}_s^t \|^2_{\mathcal{F}}
+ \underbrace{\|k\mathbf{\tilde{S}}^t\|^2_{\mathcal{F}}}_{const}
-2tr(k{\mathbf{\tilde{S}}^t}{\mathbf{B}_e^t}^T\mathbf{B}_s^t)
\\& + {\sigma}^t(\underbrace{\big\|{\mathbf{W}^t}^T\mathbf{X}_s^t\big\|^2_{\mathcal{F}}}_{const}
+ \underbrace{\big\|\mathbf{B}_s^t\big\|^2_{\mathcal{F}}}_{const}
-2tr({\mathbf{X}_s^t}^T\mathbf{W}^t\mathbf{B}_s^t))
\\ & \qquad\qquad s.t. \quad \mathbf{B}_s^t \in \{-1, 1\}^{k \times n_t},
\end{split}
\end{equation}
where the ``const" terms denote constants. The optimization problem of Eq.\ref{expand} is equivalent to
\begin{equation} \label{simplify}
\begin{split}
\min_{\mathbf{B}_s^t}\| \underbrace{{\mathbf{B}_e^t}^T\mathbf{B}_s^t}_{\text{term \Rmnum{1}}} \|^2_{\mathcal{F}}
-2tr(\underbrace{\mathbf{P}^T\mathbf{B}_s^t}_{\text{term \Rmnum{2}}})
\quad s.t.\, \mathbf{B}_s^t \in \{-1, 1\}^{k \times n_t},
\end{split}
\end{equation}
where $\mathbf{P} = k\mathbf{B}_e^t{{}\mathbf{\tilde{S}}^t}^T + {\sigma}^t{\mathbf{W}^t}^T\mathbf{X}_s^t$ and $tr(\cdot)$ is trace norm.

The problem in Eq.\ref{simplify} is NP-hard for directly optimizing the binary code matrix $\mathbf{B}_s^t$.
Inspired by the recent advance on binary code optimization \cite{shen2015supervised}, a closed-form solution for one row of $\mathbf{B}_s^t$ can be obtained while fixing all the other rows.
Therefore, we first reformulate $\text{term \Rmnum{1}}$ and $\text{term \Rmnum{2}}$ in Eq.\ref{simplify} as follows:
\begin{equation} \label{term1}
\text{term \Rmnum{1}} = {{}\mathbf{\tilde{b}}_{er}^t}^T\mathbf{\tilde{b}}_{sr}^t
+ {{}\mathbf{\tilde{B}}_e^t}^T\mathbf{\tilde{B}}_s^t,
\end{equation}
\begin{equation} \label{term2}
\text{term \Rmnum{2}} = \mathbf{\tilde{p}}_r^T\mathbf{\tilde{b}}_{sr}^t
+ \mathbf{\tilde{P}}^T\mathbf{\tilde{B}}_s^t,
\end{equation}
where $\mathbf{\tilde{b}}_{er}^t$, $\mathbf{\tilde{b}}_{sr}^t$ and $\mathbf{\tilde{p}}_r$ stand for the $r$-row of $\mathbf{B}_e^t$, $\mathbf{B}_s^t$ and $\mathbf{P}$, respectively.
Also, $\mathbf{\tilde{B}}_e^t$, $\mathbf{\tilde{B}}_s^t$ and $\mathbf{\tilde{P}}$ represent the matrix of $\mathbf{B}_e^t$ excluding $\mathbf{\tilde{b}}_{er}^t$, the matrix of $\mathbf{B}_s^t$ excluding $\mathbf{\tilde{b}}_{sr}^t$ and the matrix of $\mathbf{P}$ excluding $\mathbf{\tilde{p}}_r$, respectively.

\begin{algorithm}[t]
\caption{Balanced Similarity for Online Discrete Hashing (BSODH)}
\begin{algorithmic}[1]
\REQUIRE
    Training data set $\mathbf{X}$ with its label space $\mathbf{L}$, the number of hash bits $k$, the parameters $\sigma$ and $\lambda$, the total number of streaming data batches $T$.
\ENSURE
    Binary codes $\mathbf{B}$ for $\mathbf{X}$ and hash weights $\mathbf{W}$.\\
\FOR {$t=1 \to T$}
    \STATE Denote the newly coming data batch as $\mathbf{X}_s^t$;
    \IF{$ t = 1 $}
        \STATE Initialize $\mathbf{W}^t$ with normal Gaussian distribution;
        \STATE Compute $\mathbf{B}_s^t = sgn({{}\mathbf{W}^t}^T\mathbf{X}_s^t)$;
    \ELSE
        \STATE Compute $\mathbf{S}^t$ based on the label sets $\mathbf{L}_s^t$ and $\mathbf{L}_e^t$;
        \STATE Compute $\mathbf{\tilde{S}}^t$ via Eq.\ref{Softened_similarity};
        \STATE Initialize $\mathbf{B}_s^t = sgn({{}\mathbf{W}^t}^T\mathbf{X}_s^t)$;
        \STATE Update $\mathbf{W}^t$ via Eq.\ref{w_solution} and $\mathbf{B}_e^t$ via Eq.\ref{be_solution};
        \REPEAT
            \FOR {$r=1 \to k$}
                \STATE Update $\mathbf{\tilde{b}}_{sr}^t$ via Eq.\ref{bs_solution};
            \ENDFOR
        \UNTIL{(convergency or reaching maximum iterations)}

    \ENDIF
    \STATE Set $\mathbf{X}_e^t = [\mathbf{X}_e^t; \mathbf{X}_s^t]$ and $\mathbf{B}_e^t = [\mathbf{B}_e^t; \mathbf{B}_s^t]$;
\ENDFOR
\STATE
    Set $\mathbf{W}$ = $\mathbf{W}^t$;
\STATE
    Compute $\mathbf{B} = sgn(\mathbf{W}^T\mathbf{X})$;
\STATE
    Return $\mathbf{W}$ and $\mathbf{B}$.
\end{algorithmic}
\label{alg1}
\end{algorithm}

Taking Eq.\ref{term1} and Eq.\ref{term2} back to Eq.\ref{simplify} and expanding it, we obtain the following optimization problem:
\begin{equation}
\begin{split}
&\min_{\mathbf{\tilde{b}}_{sr}^t}
\underbrace{\|{{}\mathbf{\tilde{b}}_{er}^t}^T\mathbf{\tilde{b}}_{sr}^t\|^2_{\mathcal{F}}}_{const} +
\underbrace{\|{{}\mathbf{\tilde{B}}_e^t}^T\mathbf{\tilde{B}}_s^t\|^2_{\mathcal{F}}}_{const} +
2tr({{}\mathbf{\tilde{B}}_s^t}^T\mathbf{\tilde{B}}_e^t{{}\mathbf{\tilde{b}}_{er}^t}^T\mathbf{\tilde{b}}_{sr}^t) %
\\& - 2tr(\mathbf{\tilde{p}}_r^T\mathbf{\tilde{b}}_{sr}^t)
- 2\underbrace{tr(\mathbf{\tilde{P}}^T\mathbf{\tilde{B}}_s^t)}_{const}
\quad s.t. \quad \mathbf{\tilde{b}}^t_{sr} \in \{-1, 1\}^{n_t}.
\end{split}
\end{equation}
%

\begin{table*}[!t]
\centering
\caption{\textit{m}AP (\textit{m}AP@$1,000$) and Precision@H2 comparisons on CIFAR-$10$ and Places$205$ with hash bits of $32$, $64$ and $128$.}
\label{cifar-places-map-precisionh2}
\scalebox{0.83}[0.83]{
\begin{tabular}{c|ccc|ccc|ccc|ccc}
\hline
\multirow{3}{*}{Method}
&\multicolumn{6}{c|}{CIFAR-$10$}  &\multicolumn{6}{c}{Places$205$}  \\
\cline{2-13}
&\multicolumn{3}{c|}{\textit{m}AP} &\multicolumn{3}{c|}{Precision@H2} &\multicolumn{3}{c|}{\textit{m}AP-$1,000$} &\multicolumn{3}{c}{Precision@H2}   \\
\cline{2-13}
                        & 32-bit & 64-bit & 128-bit & 32-bit &64-bit  &128-bit & 32-bit & 64-bit & 128-bit & 32-bit &64-bit  &128-bit  \\ \hline
OKH                     &0.223   &0.268  &0.350    &0.100   &0.175   &0.372   &0.122   &0.114  &0.258    &0.026   &0.217   &0.075    \\ \hline
SketchHash              &0.302   &  -    &   -     &0.385   &   -    & -      &0.202   &  -    &  -      &0.220   &-       &-         \\ \hline
AdaptHash               &0.216   &0.305  &0.293    &0.185   &0.166   &0.164   &0.195   &0.222  &0.229    &0.012   &0.021   &0.022    \\ \hline
OSH                     &0.129   &0.127  &0.125    &0.137   &0.083   &0.038   &0.022   &0.043  &0.164    &0.012   &0.030   &0.059    \\ \hline
MIHash                  &0.675   &0.667  &0.664    &0.657   &0.500   &0.413   &0.244   &\textbf{0.308}  &0.332    &0.204   &0.202  &0.069    \\ \hline\hline
BSODH            &\textbf{0.689}&\textbf{0.709}&\textbf{0.711}&\textbf{0.691}&\textbf{0.690}&\textbf{0.602}&\textbf{0.250}&\textbf{0.308}&\textbf{0.337}&\textbf{0.241}&\textbf{0.212}&\textbf{0.101}\\
\hline
\end{tabular}}
\vspace{-1em}
\end{table*}

\begin{table}[!t]
\centering
\caption{\textit{m}AP (\textit{m}AP@$1,000$) and Precision@H2 comparisons on MNIST with hash bits of $32$, $64$ and $128$.}
\label{mnist-map-precisionh2}
\scalebox{0.83}[0.83]{\begin{tabular}{c|ccc|ccc}
\hline
\multirow{2}{*}{Method} & \multicolumn{3}{c|}{\textit{m}AP}   & \multicolumn{3}{c}{Precision@H2}          \\
\cline{2-7}
                        & 32-bit & 64-bit& 128-bit & 32-bit &64-bit & 128-bit \\ \hline
OKH                     &0.224   &0.301  &0.404    &0.457   &0.522  &0.124    \\ \hline
SketchHash              &0.348   &  -    &  -      &0.691   & -     &-        \\ \hline
AdaptHash               &0.319   &0.292  &0.208    &0.535   &0.163  &0.168    \\ \hline
OSH                     &0.130   &0.146  &0.143    &0.192   &0.109  &0.019    \\ \hline
MIHash                  &0.744   &0.713  &0.681    &0.814   &0.720  &0.471    \\ \hline\hline
BSODH            &\textbf{0.747}&\textbf{0.766}&\textbf{0.760}&\textbf{0.826}&\textbf{0.814}&\textbf{0.643}\\ \hline
\end{tabular}}
\vspace{-1em}
\end{table}

Note that $\|{{}\mathbf{\tilde{b}}_{er}^t}^T\mathbf{\tilde{b}}_{sr}^t\|^2_{\mathcal{F}} = k^2$, which is a constant value. The above optimization problem is equivalent to:
\begin{equation}
\min_{\mathbf{\tilde{b}}_{sr}^t}
tr(({{}\mathbf{\tilde{B}}_s^t}^T
\mathbf{\tilde{B}}_e^t{{}\mathbf{\tilde{b}}_{er}^t}^T -
\mathbf{\tilde{p}}_r^T)\mathbf{\tilde{b}}_{sr}^t)
\quad s.t. \quad \mathbf{\tilde{b}}^t_{sr} \in \{-1, 1\}^{n_t}.
\end{equation}

Therefore, this sub-problem can be solved by the following updating rule:
\begin{equation} \label{bs_solution}
\mathbf{\tilde{b}}_{sr}^t = sgn
(\mathbf{\tilde{p}}_r -
\mathbf{\tilde{b}}_{er}^t
{{}\mathbf{\tilde{B}}_e^t}^T
\mathbf{\tilde{B}}_s^t).
\end{equation}

The main procedures of the proposed BSODH are summarized in Alg.\ref{alg1}. Note that, in the first training stage, $\emph{i.e.}, t = 1$, we initialize $\mathbf{W}^1$ with normal Gaussian distribution as in line $4$ and compute $\mathbf{B}_s^1$ as in line $5$. When $t \geq 2$, we initialize $\mathbf{B}_s^t$ in line $9$ to fasten the training iterations from line $11$ to line $15$. By this way, it is quantitatively shown in the experiment that it takes only one or two iterations to get convergence (see Fig.\ref{iter}).

\section{Experiments}

\subsection{Datasets}
\textbf{CIFAR-$\mathbf{10}$} contains $60$K samples from $10$ classes, with each represented by a $4,096$-dimensional CNN feature \cite{simonyan2014very}.
Following \cite{fatih2017mihash}, we partition the dataset into a retrieval set with $59$K samples, and a test set with 1K samples.
From the retrieval set, $20$K instances are adopted to learn the hash functions.

\textbf{Places$\mathbf{205}$} is a $2.5$-million image set with $205$ classes.
Following \cite{fatih2017mihash,cakir2017online}, features are first extracted from the \textit{fc7} layer of the AlexNet \cite{krizhevsky2012imagenet}, and then reduced to $128$ dimensions by PCA.
$20$ instances from each category are randomly sampled to form a test set, the remaining of which are formed as a retrieval set.
$100$K samples from the retrieval set are sampled to learn hash functions.

\textbf{MNIST} consists of $70$K handwritten digit images with $10$ classes, each of which is represented by $784$ normalized original pixels.
We construct the test set by sampling $100$ instances from each class, and form a retrieval set using the rest.
A random subset of $20$K images from the retrieval set is used to learn the hash functions.

\subsection{Baselines and Evaluated Metrics}
We compare the proposed BSODH with several state-of-the-art OH methods,
including Online Kernel Hashing ($\mathbf{OKH}$) \cite{huang2013online}, Online Sketch Hashing ($\mathbf{SketchHash}$) \cite{leng2015online}, Adaptive Hashing ($\mathbf{AdaptHash}$) \cite{cakir2015adaptive}, Online Supervised Hashing ($\mathbf{OSH}$) \cite{cakir2017online} and OH with Mutual Information ($\mathbf{MIHash}$) \cite{fatih2017mihash}.

To evaluate the proposed method, we adopt a set of widely-used protocols including mean Average Precision (denoted as \emph{$\textbf{m}$}$\mathbf{AP}$),
mean precision of the top-R retrieved neighbors (denoted as $\mathbf{Precision@R})$ and precision within a Hamming ball of radius  $2$ centered on each query (denoted as $\mathbf{Precision@H2}$).
Note that, following the work of \cite{fatih2017mihash}, we only compute \emph{m}AP on the top-$1,000$ retrieved items (denoted as $\emph{\textbf{m}}\mathbf{AP@1,000}$) on Places$205$ due to its large scale.
And for SketchHash \cite{leng2015online}, the batch size has to be larger than the size of hash bits.
Thus, we only report its performance when the hash bit is $32$.

\subsection{Quantitative Results}
\begin{figure}
\begin{center}
\centerline{
\includegraphics[height=0.35\linewidth]{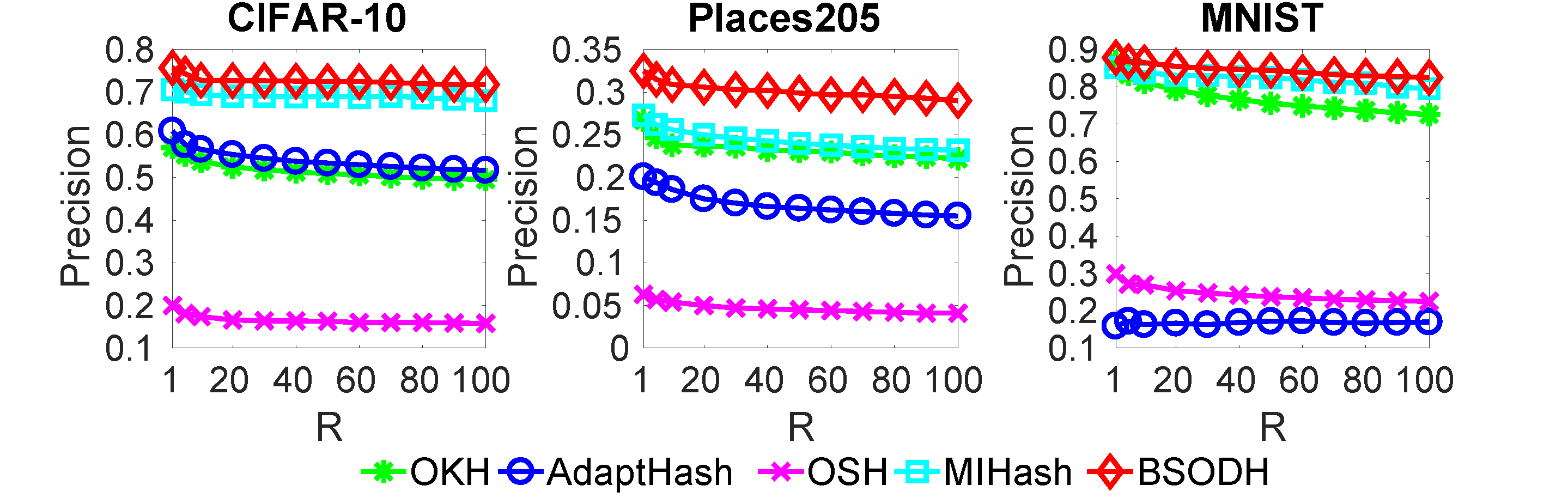}}
\caption{\label{precr}Precision@R curves of compared algorithms on three datasets with hash bit of $64$.}
\end{center}
\vspace{-1em}
\end{figure}

We first show the experimental results of \emph{m}AP (\emph{m}AP@$1,000$) and Precision@H$2$ on CIFAR-$10$, Places$205$ and MNIST.
The results are shown in Tab.\ref{cifar-places-map-precisionh2} and Tab.\ref{mnist-map-precisionh2}.
Generally, the proposed BSODH is consistently better in these two evaluated metrics on all three benchmarks.
For a depth analysis, in terms of \emph{m}AP, compared with the second best method, \emph{i.e.}, MIHash, the proposed method achieves improvements of $5.11\%$, $1.40\%$, and $6.48\%$ on CIFAR-$10$, Places-$205$ and MNIST, respectively.
As for Precision@H$2$, compared with MIHash, the proposed method acquires $29.97\%$, $2.63\%$ and $9.2\%$ gains on CIFAR-$10$, Places-$205$ and MNIST, respectively.
We also evaluate Precision@R with R ranging from $1$ to $100$ under the hash bit of $64$. The experimental results are shown in Fig.\ref{precr}, which verifies that the proposed BSODH also achieves superior performance on all three benchmarks.

\subsection{Parameter Sensitivity}
The following experiments are conducted on MNIST with the hash bit fixed to $64$.

\begin{figure*}
\begin{center}
\centerline{
\includegraphics[height=0.16\linewidth]{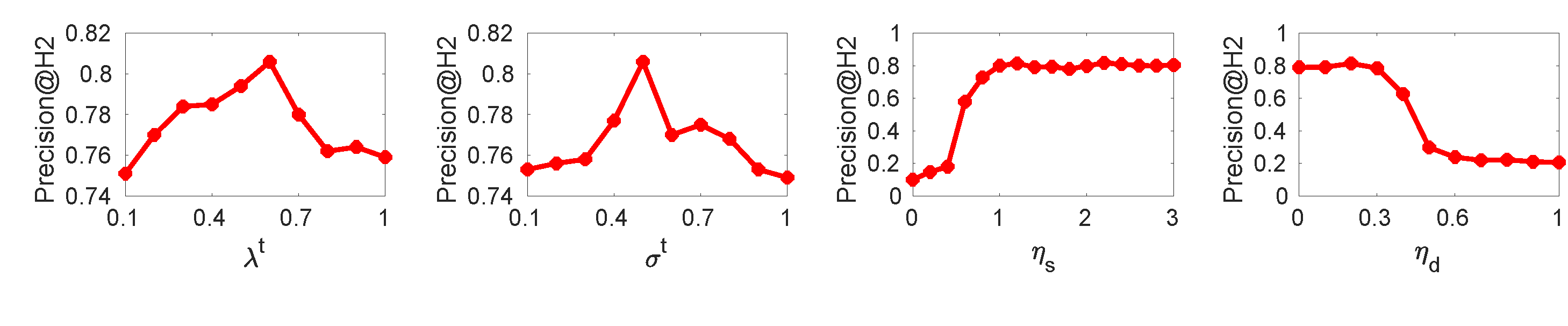}}
\caption{\label{all_parameters} Precision@H2 with respect to varying values of ${\lambda}^t$, ${\sigma}^t$, ${\eta}_s$ and ${\sigma}_d$.}
\end{center}
\vspace{-1em}
\end{figure*}

\textbf{Sensitivities to ${\lambda}^t$ and ${\sigma}^t$}.
The left two figures in Fig.\ref{all_parameters} present the effects of the hyper-parameters ${\lambda}^t$ and ${\sigma}^t$. For simplicity, we regard ${\lambda}^t$ and ${\sigma}^t$ as two constants across the whole training process.
As shown in Fig.\ref{all_parameters}, the performance of the proposed BSODH is sensitive to the values of ${\sigma}^t$ and ${\lambda}^t$.
The best combination for $({\lambda}^t, {\sigma}^t)$ is $(0.6, 0.5)$.
By conducting similar experiments on CIFAR-$10$ and Places-$205$, we finally set the tuple value of $({\lambda}^t, {\sigma}^t)$ as $(0.3, 0.5)$ and $(0.9, 0.8)$ for these two benchmarks.

\textbf{Necessity of $\mathbf{\tilde{S}}^t$}.
We validate the effectiveness of the proposed balanced similarity $\mathbf{\tilde{S}}^t$ by plotting the Precision@H$2$ curves with respect to the two positive equilibrium factors, \emph{i.e.}, ${\eta}_s$ and ${\eta}_d$.
As shown in the right two figures of Fig.\ref{all_parameters}, the performance stabilizes when ${\eta}_s \geq 1$ and ${\eta}_d \leq 0.3$.
When ${\eta}_d = 1$ and ${\eta}_s = 1$, $\mathbf{\tilde{S}}^t$ degenerates into an un-balanced version $\mathbf{S}^t$.
However, as observed from the rightmost chart in Fig.\ref{all_parameters}, when ${\eta}_s = 1$, the proposed method suffers from severe performance loss.
Precisely, the Precision@H$2$ shows the best of $0.814$ when ${\eta}_s = 1.2$ and ${\eta}_d = 0.2$, while it is only $0.206$ when ${\eta}_s = 1$ and ${\eta}_d = 1$.
Compared with the un-balanced $\mathbf{S}^t$, the proposed balanced similarity $\mathbf{\tilde{S}}^t$ gains a $295.15\%$ increase,
which effectively shows the superiority of the proposed balanced similarity $\mathbf{\tilde{S}}^t$.
In our experiment, we set the tuple $({\eta}_s, {\eta}_d)$ as $(1.2, 0.2)$ on MNIST.
Similarly, it is set as $(1.2, 0.2)$ on CIFAR-$10$ and $(1, 0)$ on Places$205$.

\begin{figure}[!t]
\begin{center}
\begin{minipage}[t]{0.33\linewidth}
\centerline{
\subfigure[$\mathbf{B}_e^t$]{
\includegraphics[width=\linewidth]{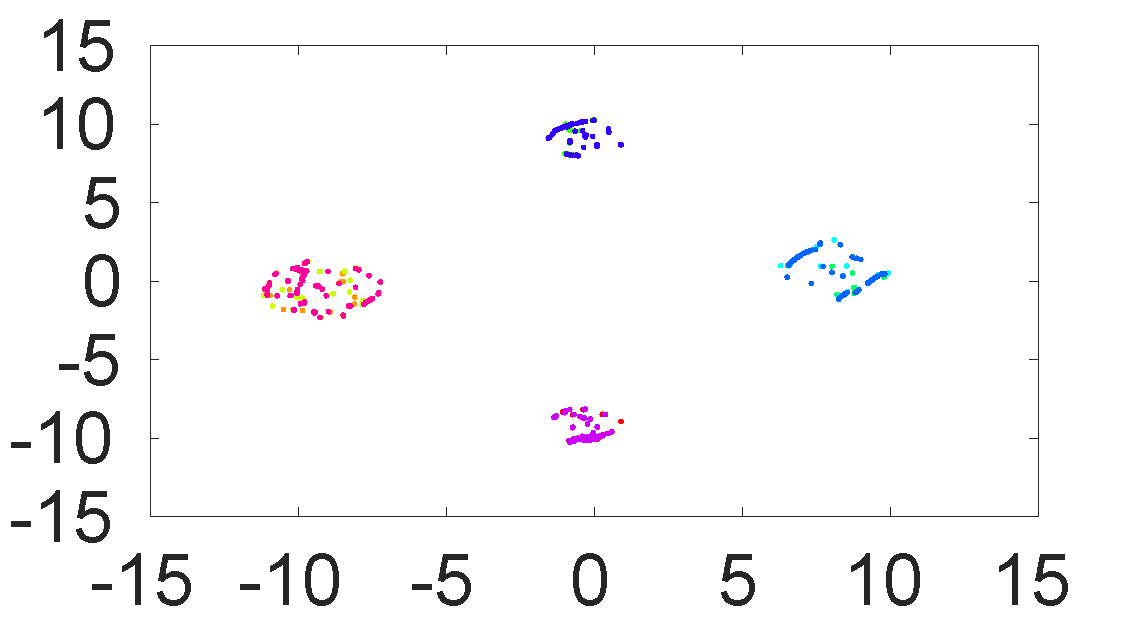}}
\subfigure[$\mathbf{B}_s^t$]{
\includegraphics[width=\linewidth]{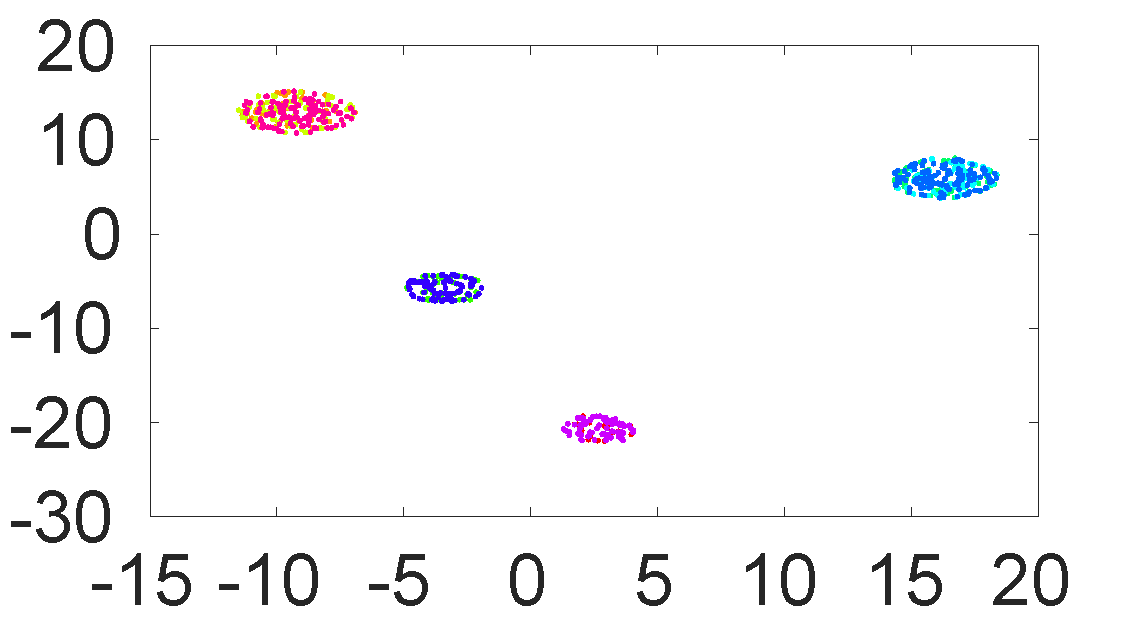}}
\subfigure[$\mathbf{sgn}({{}\mathbf\mathbf{W}^t}^T\mathbf{X}_s^t)$]{
\includegraphics[width=\linewidth]{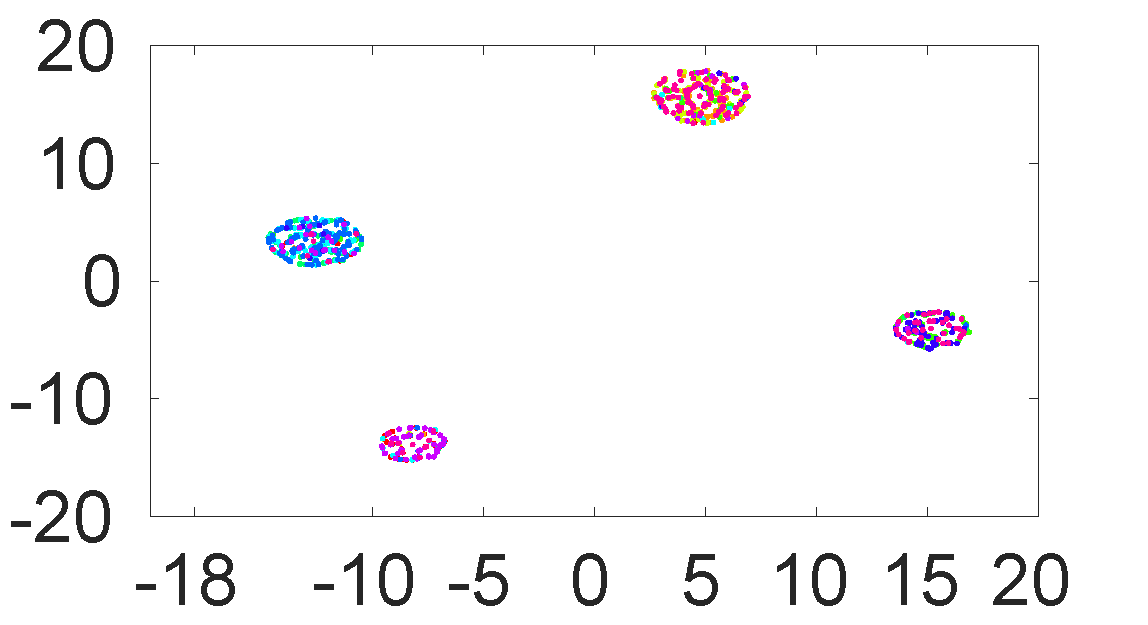}}
}
\end{minipage}

\begin{minipage}[t]{0.33\linewidth}
\centerline{
\subfigure[$\mathbf{B}_e^t$]{
\includegraphics[width=\linewidth]{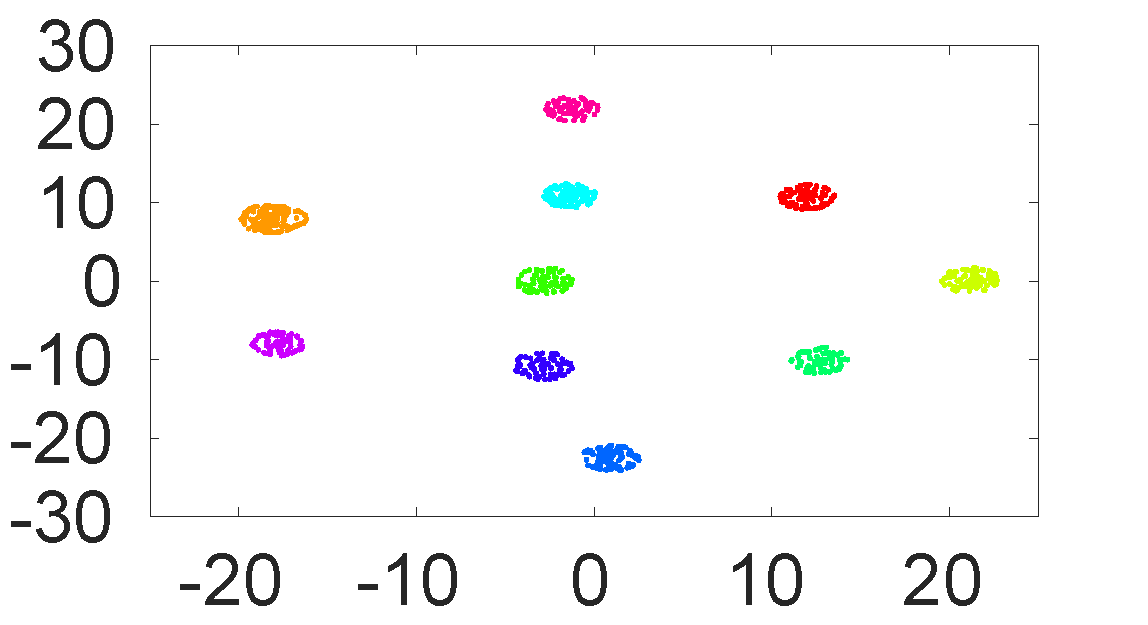}}
\subfigure[$\mathbf{B}_s^t$]{
\includegraphics[width=\linewidth]{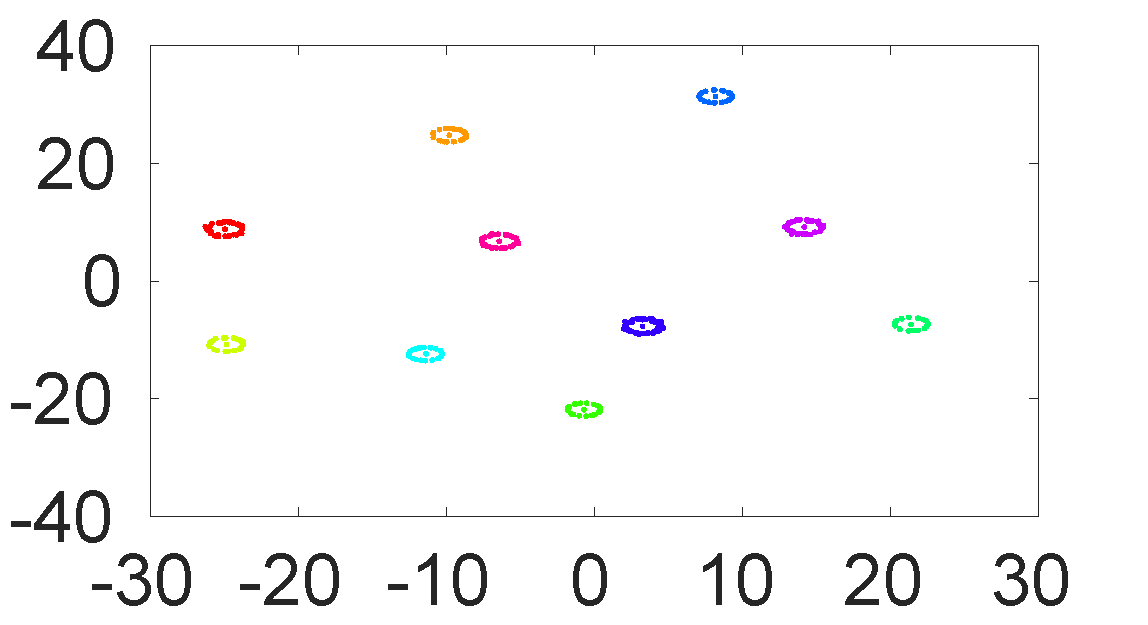}}
\subfigure[$\mathbf{sgn}({{}\mathbf\mathbf{W}^t}^T\mathbf{X}_s^t)$]{
\includegraphics[width=\linewidth]{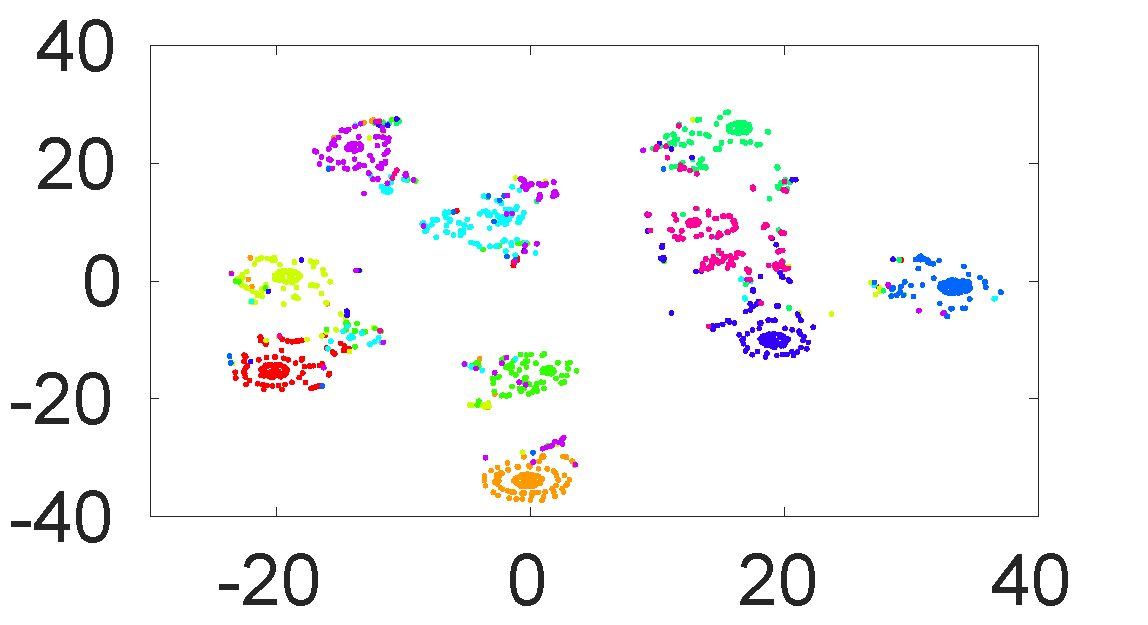}}
}
\end{minipage}

\end{center}
\caption{\label{svisual}The t-SNE visualization of hash codes. The top row shows the un-balanced results. The bottom row shows the balanced results. Given $10$ data clusters, only four are formed for un-balanced results due to the existence of data-imbalance problem. It can be solved by the proposed balanced similarity with more clusters being formed.}
\end{figure}

To verify the aforementioned \textbf{Analysis 1} and \textbf{Analysis 2},
we further visualize the learned binary codes in the last training stage via t-SNE \cite{maaten2008visualizing}. As shown in Fig.\ref{svisual},
(a), (b) and (c) are derived under un-balanced similarity $\mathbf{S}^t$ with ${\eta}_s = 1$ and ${\eta}_d = 1$. And Fig.\ref{svisual} (d), (e) and (f) are obtained under balanced similarity $\mathbf{\tilde{S}}^t$ with ${\eta}_s = 1.2$ and ${\eta}_d = 0.2$.
%

%
Though the discretely optimized binary codes $\mathbf{B}_e^t$ (a), $\mathbf{B}_s^t$ (b) and linearly mapped binary codes $sgn({{}\mathbf{W}^t}^T\mathbf{X}_s^t)$ (c) are clustered, each cluster is mixed with items from different classes and only four out of ten clusters are formed with each close to each other.
That is to say, the majorities of Hamming codes are the same, which conforms with \textbf{Analysis 1}.
However, under the balanced setting, both $\mathbf{B}_e^t$ and $\mathbf{B}_s^t$ are formed into ten separated clusters without mixed items in each clusters, which conforms with \textbf{Analysis 2}.
Under such a situation, the hash functions $\mathbf{W}^t$ are well deduced by $\mathbf{B}_s^t$, with the hash codes in Fig.\ref{svisual} (f) more discriminative.

\begin{figure}
\begin{center}
\centerline{
\includegraphics[height=0.35\linewidth]{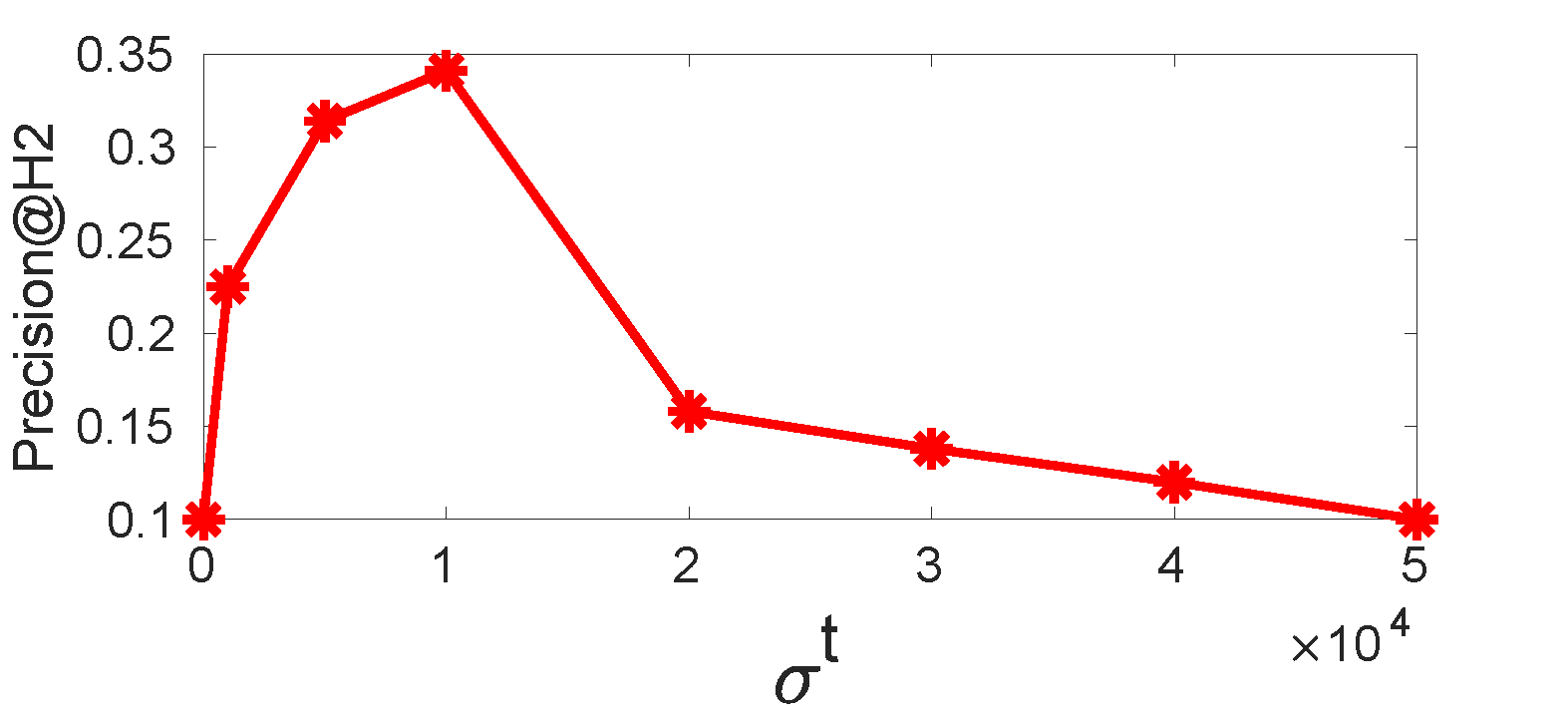}}
\caption{\label{sigma} Precision@H2 results when scaling up ${\sigma}^t$.}
\end{center}
\vspace{-1em}
\end{figure}

\textbf{Scaling up ${\sigma}^t$}.
As aforementioned, an alternative approach to solving the data-imbalance problem in \textbf{Analysis 1} is to keep a balance between term $1$ and term $2$ in Eq.\ref{original_binary_independent} via scaling up the parameter ${\sigma}^t$.
To test the feasibility of this scheme, we plot the values of Precision@H2 with ${\sigma}^t$ varying in a large scale in Fig.\ref{sigma}.
Intuitively, scaling up ${\sigma}^t$ affects the performance quite a lot. Quantitatively, when the value of ${\sigma}^t$ is set as $10,000$, Precision@H$2$ achieves the best, \emph{i.e.,} $0.341$.
We argue that this scheme shows its drawbacks in two aspects.
First, it suffers from the unsatisfactory performance. As shown in Tab.\ref{mnist-map-precisionh2}, when hash bit is $64$, the proposed BSODH gets $0.814$ in term of Precision@H$2$ on MNIST. Compared with scaling up ${\sigma}^t$, the proposed method achieves more than $2.5$ times better performance.
Second, scaling up ${\sigma}^t$ also easily gets stuck in how to choose an appropriate value due to the large range of ${\sigma}^t$. To decide a best value, extensive experiments have to be repeated, which is infeasible in online learning. However, ${\sigma}^t$ is limited to $[0, 1]$ under the proposed BSODH. It is much convenient to choose an appropriate value for ${\sigma}^t$.

\begin{figure}
\begin{center}
\centerline{
\includegraphics[height=0.33\linewidth]{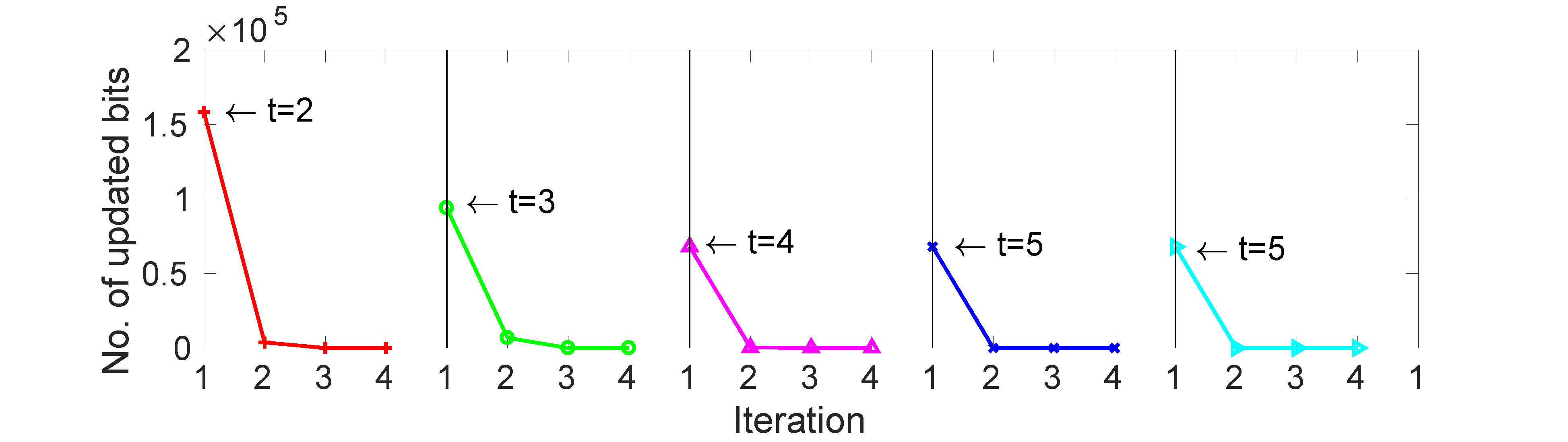}}
\caption{\label{iter} Convergence of the proposed BSODH.}
\end{center}
\end{figure}

\textbf{Convergence of $\mathbf{B}_s^t$}.
Each time when the new streaming data arrives, $\mathbf{B}_s^t$ is updated based on iterative process, as shown in lines $11-15$ in Alg.\ref{alg1}.
Fig.\ref{iter} shows the convergence ability of the proposed BSODH on the input streaming data at $t$-stage. As can be seen, when $t \leq 2$, it merely takes two iterations to get convergence. What's more, it costs only one iteration to finish updating $\mathbf{B}_s^t$ when $t > 2$, which validates not only the convergence ability, but also the efficiency of the proposed BSODH.

\section{Conclusions}
In this paper, we present a novel supervised OH method, termed BSODH.
The proposed BSODH learns the correlation of binary codes between the newly streaming data and the existing database via a discrete optimization, which is the first to the best of our knowledge.
To this end, first we use an asymmetric graph regularization to preserve the similarity in the produced Hamming space.
Then, to reduce the quantization error, we mathematically formulate the optimization problem and derive the discrete optimal solutions.
Finally, to solve the data-imbalance problem, we propose a balanced similarity, where two equilibrium factors are introduced to balance the similar/dissimilar weights.
Extensive experiments on three benchmarks demonstrate that our approach merits in both effectiveness and efficiency over several state-of-the-art OH methods.

\section{Acknowledge}
This work is supported by the National Key R\&D Program (No. 2017YFC0113000, and No. 2016YFB1001503), Nature Science Foundation of China (No. U1705262, No. 61772443, and No. 61572410), Post Doctoral Innovative Talent Support Program under Grant BX201600094, China Post-Doctoral Science Foundation under Grant 2017M612134, Scientific Research Project of National Language Committee of China (Grant No. YB135-49), and Nature Science Foundation of Fujian Province, China (No. 2017J01125 and No. 2018J01106).

\bibliographystyle{aaai}
\bibliography{mybib}

\end{document}